# RECENT ADVANCEMENT IN NEXT-GENERATION SEQUENCING TECHNIQUES AND ITS COMPUTATIONAL ANALYSIS

KHALID RAZA

*Department of Computer Science, Jamia Millia Islamia, New Delhi, India*
*kraza@jmi.ac.in*

SABAHUDDIN AHMAD

*Department of Computer Science, Jamia Millia Islamia, New Delhi, India*
*sabahuddin@live.in*

Next Generation Sequencing (NGS), a recently evolved technology, have served a lot in the research and development sector of our society. This novel approach is a newbie and has critical advantages over the traditional Capillary Electrophoresis (CE) based Sanger Sequencing. The advancement of NGS has led to numerous important discoveries, which could have been costlier and time taking in case of traditional CE based Sanger sequencing. NGS methods are highly parallelized enabling to sequence thousands to millions of molecules simultaneously. This technology results into huge amount of data, which need to be analysed to conclude valuable information. Specific data analysis algorithms are written for specific task to be performed. The algorithms in group, act as a tool in analysing the NGS data. Analysis of NGS data unravels important clues in quest for the treatment of various life-threatening diseases; improved crop varieties and other related scientific problems related to human welfare. In this review, an effort was made to address basic background of NGS technologies, possible applications, computational approaches and tools involved in NGS data analysis, future opportunities and challenges in the area.

*Biography*: **Khalid Raza** is currently working as an Assistant Professor at the Department of Computer Science, Jamia Millia Islamia, New Delhi, India. He has contributed over 25 research articles in refereed international journals, conference proceedings and as book chapters. His research interests are systems biology, soft computing techniques and microarray and NGS analysis. He has received grants for two Govt. funded research projects and worked as Principal Investigator.

**Sabahuddin Ahmad** has recently joined as Research Fellow at Heinrich Heine Universität Düsseldorf, Germany. Earlier he worked as Junior Research Fellow in Central Drug Research Institute, Lucknow, India. His area of interest is NGS data analysis, molecular docking and simulation.

## 1. Introduction

DNA sequencing is a method to obtain the exact order of occurrence of nucleotides in a DNA. With the help of DNA sequence, researchers can illuminate genetic information from a biological system. Deciphering DNA sequences is necessary for almost all branches of life sciences and its understanding has grown exponentially in the past decades. The first-generation sequencing method, called Sanger sequencing, was developed by Edward Sanger in 1975. Since its inception, Sanger sequencing was considered as the gold standard for DNA sequencing for around three decades (Sanger et al., 1977). The first major breakthrough of first-generation sequencing was the Human Genome Project (HGP), 13-years long, $3 million project, completed in 2003. Due to inherent limitations in throughput, speed, scalability and resolution of first-generation Sanger sequencing approach, second-generation of sequencing method, or Next-Generation Sequencing (NGS) has been developed to cater high demand for cheaper as well as faster sequencing technology.

The NGS is fundamentally different approach to sequencing that lead to several ground-breaking discoveries and brought a new revolution in genomic research by revealing limitless insight related to genome, transcriptome and epigenome of any species. Hence, NGS technology has brought a new revolution in the welfare of human society. Principally, basic idea behind NGS is similar to capillary electrophoresis (CE) based Sanger sequencing, but NGS extends the idea to perform massive parallel sequencing, where millions of fragments of DNA from a single sample are accurately sequenced. The NGS enables sequencing of large stretch of DNA base pairs, producing hundreds of gigabases of data in a single sequential run. The NGS, also known as Massive Parallel Sequencing, allows the complete genome (of human) to be sequenced in less than a day. More recently, third generation sequencing (TGS) has evolved, however, it is still in its infancy (Hayden, 2009). Because of its maturity and understanding, among all the three sequencing generations (Fig. 1), the NGS is more widely accepted. Therefore, presented review primarily discusses the recent developments in NGS technology and related computational analysis.

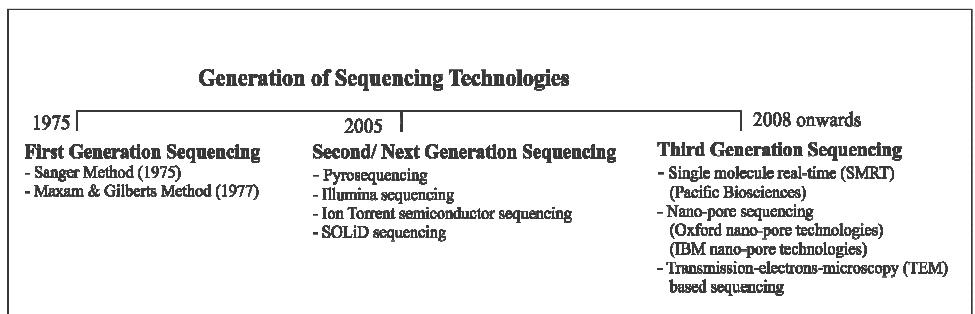

Fig. 1 Overview of sequencing generations

## 1.1. *What does NGS offer?*

The NGS is a powerful, flexibly, indispensable and universal biological tool that infused several areas of biological study. Some of the promises and capabilities of NGS are as follows:

- It provides us a much cheaper and faster alternative to traditional Sanger sequencing. Researchers can now sequence whole small genome in a day. More than five human genomes can be sequenced in a single run, generating data in less than a week, costing less than $5000 per genome (Illumina).
- High-throughput sequencing (HTS) of the human genome lets us discover genes and regulatory pathways associated with disease (Grada &Weinbrecht, 2013).
- Targeted sequencing of specific genes or genomic region helps in the identification of disease-causing mutations. It helps in the faster diagnosis, and outcome of disease-targeted sequencing may help in better therapeutic decision-making for several genetic diseases, including many cancers (Grada &Weinbrecht, 2013).
- RNA-Seq (NGS of RNA) provides entire transcriptomic information of a sample without any need of previous knowledge related to genetic sequence of an organism. RNA-Seq provides a strong alternative approach to Microarrays for gene expression studies, and let the researchers visualize RNA expression in the form of sequence (Grada &Weinbrecht, 2013).
- Variant study is common in medical genetic, where DNA sequence and data are compared with a reference sequence to catalogue the differences. These differences may range from single nucleotide polymorphisms (SNPs) to complex chromosomal rearrangement (Nekrutenko& Taylor, 2012).

## 1.2. *Evolution of NGS*

The root of the development of NGS goes back to the discovery of double helix structure of DNA by James Watson and Francis Crick in 1953. Robert Holley was the first who sequenced nucleic acid during 1964-1965. In 1977, Frederick Sanger and Walter Gilbert developed sequencing methods for long length of DNA. For these three pioneer works, the Nobel Prize was given to Watson and Crick in 1962, Holley in 1968 and Sanger and Gilbert in 1980.

The development of NGS took place in late 20th and early 21st century. The first NGS technology, known as Massively Parallel Signature Sequencing (MPSS), was launched by a USA based Lynx Therapeutics Company in 2000. This company was later taken over by Illumina – one of the big players in sequencing market. After MPSS, 454 Life Sciences launched a similar kind of sequencing technology in 2004 that gained 6-fold cost reduction compared to traditional Sanger sequencing. Later, Life Sciences was acquired by Roche Company. During 2005-2006, the 454 GS 20 Roche platform was launched that could produce 20 Mbp; and in 2007, this model was replaced by GS FLX capable of producing 100 Mbp, which was further enhanced to 400 Mbp in 2008. The Roche 454 GS-FLX was upgraded to 454 GS-FLX + Titanium sequencing platform that can produce over 600 Mbp and read length of up to 1000 bp. Roche also produced 'GS Junior' platform which handy and can be placed on a bench (Barba et al., 2014).

In 2006, Genome Analyzer was released by Solexa, which was acquired by Illumina in 2007. Solexa sequencer was capable to sequence 1 Gbp in single run (Illumina). During the last five years, Illumina has developed Mi-Seq and Hi-Seq platforms. The Mi-Seq platform can sequence up to 15 Gbp, while Hi-Seq systems achieves up to 600 Gbp. Hi-Seq 2500 system can sequence a human genome in a day. The first SOLiD sequencing system was launched in 2007, followed by SOLiD 5500w and 5500 lw in 2010. The recent SOLiD4 platform can sequence between 35-50 Gbp with a maximum read length between 35-50 bases.

Life Technologies Ion Torrent sequencer was released in 2011 that uses sequencing-by-synthesis and emulsion PCR. It does not use florescence, like other sequencing methods. This technology measures H+ ion release during base incorporation (Ion Torrent). Some of its models are PGM 314, PGM 316, PGM 318, PI and PII. The PII is launched in early 2015 with average reading length of 100 bases and sequencing capacity of up to 64 Gbp.

Complete Genomics platform is unique in the NGS market which offers the NGS services, rather than selling instrumentation. Their focus is on human whole-genome resequencing. Currently, they do not offer any other applications such as exome, RNA-Seq, ChIP-Seq, etc. Complete Genomics platform can achieve a sequencing rate of 3000 Gbp per run with maximum reading length of 70 bases. Complete Genomics is going to launch its first commercial product, the Revelocity system, which have been designed to provide total end-to-end genomic solution (Complete Genomics).

### 1.3. *NGS versus Sanger sequencing*

The traditional automated Sanger sequencing method, also called first generation sequencing, had led to several most important accomplishments that include completion of the HGP and many other animal and plant genomes. However, due to its heavy cost and time intensive task, in the late 20th and early 21st century, new methods (NGS) have been developed to replace the automated Sanger method. The major advancement offered by NGS is: i) capability to generate massive volume of data, ii) read more than one billion short reads in single run, iii) a fast and inexpensive method to get accurate genomic information.

Preparation of sample for NGS are faster and straightforward in comparison to capillary electrophoresis (CE)-based Sanger sequencing. In NGS, it may be started directly from a gDNA or cDNA library. The DNA fragments are then ligated to platform-specific oligonucleotide adapters and needs less than 1.5 hours to complete. On the other hand, CE-based Sanger sequencing needs genomic DNA that are fragmented and cloned into either bacterial or yeast artificial chromosomes (BACs/YACs). Further, each BAC/YAC are subcloned into a sequencing vector and transformed to appropriate microbial host. Before sequencing, template DNA is purified from individual colonies. These processes may consume days or even weeks, depending on genomic size. A comparative summary for sample preparation of whole genome sequencing (WGS) and targeted sequence using CE-based Sanger sequencing and NGS are shown in Table 1.

Table 1. Sample preparation in CE-based Sanger sequencing and NGS methods

| Applications | CE-based Sanger sequencing | Next-Generation Sequencing |
|---|---|---|
| Whole genome sequencing | Library preparation is more involved. Each sample must contain a single template, demanding purification from single bacterial, yeast colonies, or phage plaques from single PCR. | Library preparation is more streamlined, i.e., sample can consist of a population of DNA molecules which do not need clonal purification |
| | Sample preparation is time consuming, taking days to weeks dependent on genome size. | It is accomplished within hours, regardless of genome size. |
| Targeted sequencing | Library preparation is again more involved. | Library preparation is more streamlined. |
| | Suitable for sequencing amplicons and clone checking. | Suitable for sequencing amplicons and clone checking |
| | This process takes days to weeks, depending upon the genome size. | Completed within hours |

### 1.4. *NGS versus Microarray technology*

The measurement of mRNA expression under different experimental conditions has been of interest to the biologist. In the past decade, microarray technology has been extensively applied to simultaneously monitor gene expression on a genome-wide scale (Harismendy et al., 2009). In microarray technology, cDNA is hybridized on an array of complementary oligonucleotide probes corresponding to gene sunder study, and the abundance of mRNA is completed from its hybridization intensity. A detail discussion on microarray platforms and techniques can be found in (Allison et al., 2006; Ness, 2007) and analysis of microarray data can be found in (Quackenbush, 2001; Raza and Parveen, 2012a; Raza and Parveen, 2012b; Raza and Parveen, 2013; Raza and Jaiswal, 2013; Raza, 2014; Raza and Hasan, 2015; Raza and Kohli, 2015; Raza,2015; Raza, 2016). Microarray technology has been widely used for the study of cancer genomes and transcriptome but some of its promises have not materialized due to limitations of microarray data. For instance, microarray based gene expression profiles only give a semi-quantitative assessment of gene expression due to nature of the probes included in the platform. Also, microarray technology is unable to give any information related to structural genomic aberrations and bases pair mutations (Reis-filho, 2009).

The NGS has replaced the limitation of microarrays to detect poorly expressed genes. In 2008, Nature News published an article "The death of microarrays?' stating that "High-throughput gene sequencing seems to be stealing a march on microarrays" and "Sequencing does put microarrays at risk in some areas of the life-sciences research market" (Ledford, 2008).NGS is used as a tool for the identification and analysis of DNA regions which interact with regulatory proteins in functional regulation of gene expression. It offers novel, and rapid means for genome-wide characterisation and profiling of mRNAs, small RNAs, transcription factor regions, chromatin structure and DNA methylation patterns (Ansorge, 2009).

### 1.5. *Challenges in data interpretation*

The 'Big Data Problem' is a pressing issue originated with social networking, life sciences and bioinformatics research. The real difficulty with biomedical research is lack of appropriate computational tool and infrastructure to carry out analysis and interpretation of modern vast datasets (such as NGS data). Some of the major issues with analysis, interpretation, reproducibility and accessibility of NGS data have been pointed out by Nekrutenko and Taylor (2012). Some of these issues, extracted from Nekrutenko and Taylor (2012), are described as follows:

- *Adoption of existing analysis practices:* There must be some well-defined and accepted data analysis practices so that it can be adopted as common analysis practice. For instance, 1000 Genome Project (GPC, 2010) and HapMap (Gibbs et al., 2003) are highly coordinated effort that defines series of accepted practices for 'variant discovery'. Nekrutenko and Taylor (2012) found that despite of well documented analytical procedure developed by 1000 Genome Project, the research communities are still using a mix of heterogeneous approaches for various studies.
- *Difficulty of reproducibility:* Research publications would be influential, if it maintains complete detail on how to repeat the reported analysis so that other researchers can adopt similar approaches in their own research studies. Nekrutenko and Taylor (2012) reported that today's publications reporting computational analysis lack to provide sufficient information, such as input data sets, program source code or scripts, all parameter settings so that computational reproducibility may be achieved. Similarly, NGS data analysis results cannot be accurately verified, reproduced, adopted or used, that creates a shocking reproducibility crisis (Nekrutenko&Taylor, 2012).
- *Potential of integrative resources:* As described in (Nekrutenko& Taylor, 2012), most biological researchers face difficulties in computationally complex analysis on large datasets, and also computation analysis steps are not well documented. In this direction, a number of integrative frameworks have been developed which bring diverse tools together under the aegis of unified interface. Some of these frameworks are BioExtract (Lushbough, 2010), Galaxy (Goecks et al., 2010), GenePattern (Reich et al., 2006), GeneProf (Halbritter et al., 2011), Mobyle (Néron, et al., 2009) and so on. These potential integrative resources empower bioinformaticians and biomedical researchers to make use of powerful computing infrastructure, required for NGS data analysis.
- *Making HPC infrastructure useful to biologists:* Analysing NGS data is computation extensive. There are several powerful high performance computing (HPC) resources that may be applied for NGS computation. The HPC resources may include computing clusters that can be distributed over different geographical regions. With the popularity of Cloud Computing, many vendors are now offering 'software as a service' analysis solution for sequence data, including DNAnexus[a] and GenomeQuest[b]. There are some free and open-source software (FOSS) that can be scaled in cloud infrastructure, but these are basically single purpose. Some of the FOSS tools are Crossbow (Langmead et al., 2009) used for variant discovery, Myrna (Langmead& Hansen, 2010) for RNA-Seq analysis and CloVR (Angiuoli et al., 2011) for metagenomic annotations.

- *Improving long-term archiving:* One of the major concerns for analysis services is its longevity, i.e., an online analysis tool is not guaranteed to be online forever. One promising solution is to archive the snapshots of a particular analysis (Nekrutenko& Taylor, 2012).

### 1.6. *Limitations of NGS and promises of TGS*

- Although, NGS is cheaper and faster in comparison to traditional Sanger sequencing but still it is too expensive to be affordable by small labs or an individual.
- NGS data analysis is time-consuming and needs sufficient knowledge of bioinformatics to harvest accurate information from these sequence data.
- The National Human Genome Research Institute (NHGRI) posed an open goal to minimize the cost of human genome sequencing to $1000 to fuel further development in NGS technologies so that routine sequencing of human genomes can be used as a tool to diagnose various diseases. But, till date this target is still too far from its goal of $1000 genome (Hert et al., 2008).
- NGS supports read lengths of small size, which results into highly repetitive sequences. Support for sequencing from short read lengths is one of the major shortcoming which limit its application, especially in de novo sequencing (Hert et al., 2008).
- Data processing steps is another major bottleneck for the implementation and capitalization of NGS technology (Daber et al., 2013).

Third generation sequencing (TGS) takes advantage of the above gaps of NGS technologies and allows direct sequencing of single molecules. Single molecule sequencing (SMS) technologies have the ability to sequence with longer read lengths keep the cost and time both lower without compromising the quality of the genome assembly.

### 2. Methodology of Next Generation Sequencing

### 2.1. *How the stuff works*

Basic principle on which NGS works is similar to traditional Sanger sequencing methods involving capillary electrophoresis. Mostly, different NGS platforms adopt their own specific protocol in the sequencing methods. NGS starts with template preparation, and generally, the double-stranded DNA is considered as the starting material. However, the source from which this material is derived may vary. Sources can be either genomic DNA, immuno-precipitated DNA, reverse-transcribed RNA or cDNA (Rizzo et. al., 2012). The next step involves library preparation. If genomic DNA was considered in the previous step, it should be fragmented into linear DNA molecules. If RNA was considered in the previous step by the means of reverse transcription, cDNA is obtained for preparation of linear DNA molecules. Moreover, the fragmentation of these DNA molecules into smaller fragments is involved, which along with specific size selection steps serve to break the considered DNA template into smaller, but sequence-able fragments depending upon requisite platform. Moreover, adapter ligation is also involved in this process, which adds platform specific, synthetic DNAs at the end of the DNA fragments present in this library to facilitate the sequencing reactions. Thus, sequence library preparation involves some common steps of fragmentation, size selection and adapter ligation.

---

[a]http://www.dnanexus.com
[b]http://genomequest.com/

Next step involves library amplification, to produce significant signal for nucleotide addition. This step involves either by the attachment of DNA fragment to microbead or attachment of the same to glass slide, when some PCR techniques are followed. Library amplification eventually leads into the sequencing reaction and imaging process. Ultimate destiny of NGS is to analyse the data produced through this sequencing process. NGS data analysis is complex and crucial task. It involves assessment of some important and vital genes or regulatory elements in the given genome. There are several tools available which can be either downloaded to local machine or used on the Web for NGS data analysis. A short description of the workflow for NGS is depicted in the Fig 2.

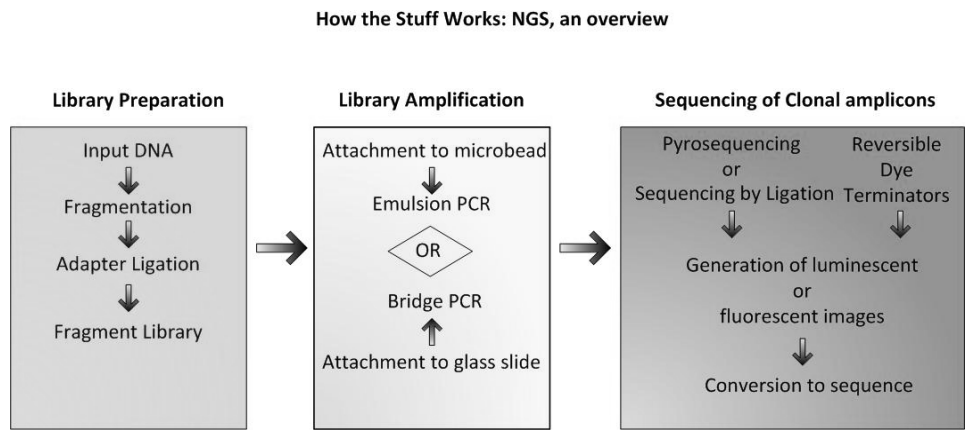

Fig. 2 How stuff works: NGS an overview

## 2.2. *Major platforms in NGS*

*Illumina Sequencing* is one of the major platforms used in NGS. In this sequencing technology reads of 100-150bp are considered. Comparatively longer fragments are considered from the template library for ligation using generic adaptors. Further, these ligated fragments are annealed (attached) to a glass slide. In the next stage, Polymerase Chain Reaction (PCR) is carried out, in order to amplify and make several copies of a read. Later, amplified copies of reads are separated into single strands, which is to be sequenced. Next, the slide is flooded with fluorescently labelled nucleotides (A, G, C & T) and DNA polymerase. Each nucleotide has a unique colour to identify itself. Terminator ensures that only one base is added at a time. The fluorescent signal in form of image is recorded, which indicates the base added. At this stage, glass slide is re-prepared for next cycle. Terminators are removed, which allows the next base to add in this process. At the same time, fluorescent signal is also removed in order to prevent future contaminations of the signals. The process is repeated by addition of a nucleotide at a time and imaging process in between. Using the computers, the bases are detected at each site in each image, which in turn help in construction of a sequence.

*454 sequencing technology* has the ability to sequence much longer reads as compared to Illumina. The basic principle is similar to Illumina, where the optical signals are read on the addition of the bases. Though 454 sequencing involves fragmentation of DNA or RNA too, in this case, the fragmentation results into 1kb long reads. In the process of library amplification, the fragments are attached to the microbeads. Using PCR, the fragments are amplified. Here, the wells also contain DNA polymerase and sequencing buffers. In the next stage, the slide is flooded with any one of the four nucleotide species (A, G, C, or T). Addition of each nucleotide releases an optic signal. The locations of signals are spotted to identify the bead where the nucleotides are added to. The process recycles after washing the mix. 454 sequencing generates graphs for each sequence read, which reflects the signal density for each nucleotide wash. Using computational approach, the density of signals in each wash facilitates the retrieval of the sequence.

*Ion Torrent PGM* (Personal Genome Machine) sequencing is different from Illumina and 454 sequencing platforms. This sequencing platform does not uses optic signals. They utilize the concept that the addition of a dNTP to a DNA polymer releases an H+ ion. Here template DNA or RNA is fragmented into size of ~200bp. Using emulsion PCR, the amplification takes place. Just as in case 454 sequencing, the slide is flooded with single types of dNTP, along with buffers and polymerase. H+ ion released on addition of dNTP to a DNA polymer decreases the pH. The pH changes are detected and recorded from each well, which allows determining the bases type and its concentration in that well.

*Helicos (HeliScope Genetic Analysis System)* is the first commercial NGS platform which used the concept of single molecule fluorescent sequencing. Single molecule sequencing strategy process simplifies the DNA sample preparation process and avoids PCR induced errors. In this process, DNA is simply trimmed into smaller fragments, tailed with poly A, and hybridized to a flow cell surface containing oligo-dT for sequencing-by-synthesis of billions of molecules in parallel (Thompson et. al., 2010). Requirement for less material adds a plus to this sequencing technique as compared to other technologies.

*Pacific Biosciences's SMRT* is single molecule real-time sequencing technologies which is popular from the last few years due to its higher accuracy, suitability for small genome, and study of facile closure bacterial genome without any additional experiment (van Dijk et al., 2014). It provides considerably longer and highly accurate DNA sequence from single molecule and deciphers where methylated bases occur, which is usually not produced by other sequencing platforms (Roberts et al., 2013). The average reads of current Pacific Biosciences's instrument is 10 Kbp or longer, which is almost 100 times longer than other NGS platforms, 100 times smaller amount of gaps compared to NGS (Sakai et al., 2015). In several studies it has been revealed that SMRT-based assemblies facilitate a more comprehensive gene annotation than NGS-based assemblies where lots of genes are generally missed or fragmented (Sakai et al., 2015). The SMRT sequencing technologies are considered as third generation sequencing (TGS). Some of the applications of SMRT are *de nova* assembly of novel genomes, DNA methyltransferase, etc. A comparative view of different sequencing platform and their sequencing chemistry, template preparation, strength and weaknesses is shown in Table 2.

Table 2. A comparative reviews of various NGS and TGS platforms

| Sequencing methods | Sequencing platform | Sequencing chemistry | Template preparation | Strength | Limitation |
|---|---|---|---|---|---|
| Sequence by synthesis | Illumina | reverse terminator | solid-phase | widely used; large throughput[a]; short run time[b] | low number of total reads[b] |
| Pyrosequencing | Roch 454 | Pyrosequencing | emulsion PCR | longer reads; short run time | reagents are costly |
| Ion semiconductor | Ion torrent | $H^+$ ion sensitive transistor | emulsion PCR | low cost; short run time | low number of total reads |
| Single molecule florescent sequencing | Helicos | reverse terminator | single molecule | free from errors occurring from amplification process | not commonly used; company provides sequencing services |
| Single molecule real-time (SMRT) | Pacific Biosciences | sequence by synthesis | single molecule | Faster and longer read lengths, allows detection of nucleotide modification | very expensive and needs huge storage and computation power |

## 3. Application Areas

The NGS has served in several aspects of medical and biochemical researches worldwide. Below we discuss some widely used applications of NGS.

### 3.1. *Whole genome sequencing (WGS)*

The WGS involves sequencing of entire genome, for instance HGP is a good example of whole genome sequence which was completed in 2013. HGP was based on traditional capillary electrophoresis-base Sanger sequencing method that cost ~3 billion USD and was completed in more than ten years. In contrast, NGS performs massively parallel sequencing which facilitates the HTS and thereby responsible for entire genome sequencing in less than a day at minimum cost. Human genome sequencing helps in identifying the specific genes and their regulatory partners involved in the pathological processes (Grada et. al., 2013). The WGS provides deeper understanding for genetic variations (impacting the normal function of a body), drug response on a gene set, and many other complex/simple biological processes. In the coming days, there will be more favourable applications of WGS for human welfare.

### 3.2. *Exome sequencing*

It helps in investigating the protein coding regions within the genome. Exome shares only small portion of the whole genome. Exomes are gene-coding region and therefore important site of concern for clinicians and researchers. Mutations, i.e., alterations in this part of genome results are possible causes of a disease or other abnormal conditions. In case of human genome, exome represents less than 2% of the total genome, but contains nearly 85% of disease-causing variants. Therefore, sequencing this part of genome in-spite of WGS results in the cost-effective sequencing. Important applications of exome sequencing involve variant association studies, discovering relevant mutations in exome region, identification of disease-related variations and so on.

### 3.3. *Targeted resequencing*

It involves in the sequencing of genomic regions of interest. This region is generally a small subset of the genome, such as exome, a set of genes, a particular chromosome or a region of particular interest. This enables in reduced cost sequencing approach. Moreover, narrowing down the search region enables one to get closer variation analysis which could have been impossible, when considering genome analysis on broader terms. Targeted resequencing has been found useful in discovering the point mutations, insertion or deletions, gene rearrangements and variations occurring in copy number.

### 3.4. *Chromatin immunoprecipitation sequencing (ChIP-Seq)*

It is used to analyse the interactions of proteins with DNA and RNA. ChIP-Seq combines the ChIP technique and NGS technique to identify the binding sites of the DNA associated proteins. This enables in interpreting the regulation events which may be causative to many biological processes such as gene regulation, DNA repair, and DNA synthesis. ChIP-Seq also serves important role in classifying the disease situations resulting from abnormal functioning of the genomic contents and their expression levels. Important applications of ChIP-Seq include analysis of epigenetic modifications, mapping of epigenetically modified DNA sequences, interactions of viral protein with human genome resulting in complex disease situations.

### 3.5. *RNA sequencing (RNA-Seq)*

RNA sequencing (RNA-Seq) is a transcriptome sequencing approach and involves a wide variety of applications, ranging from simple mRNA profiling to discovery of the entire transcriptome. RNA-Seq is being popular nowadays because they are found uncovering information which may have been missed by array-based platforms regardless of prior knowledge of the transcript sequence. Key features of RNA-Seq include: detection of transcripts with low expression levels, and transcript analysis with or without reference sequence. Moreover, RNA-Seq is found to be a powerful tool for accurate quantification of expression levels.

The above list of applications of NGS is not exhaustive. There are several other applications directly or indirectly attached with HTS. Moreover, as stated before, NGS is a newbie, but with the passage of time there will be several improvements and additions to this technology which shall help for quality researches for human welfare.

## 4. Database Archives

NGS platforms are producing unprecedented amount of sequence data. Some of the popular and commonly used NGS database archives are as follows:

### 4.1. *Sequence Read Archive*

To archive large volumes of NGS data, National Centre for Biotechnology Information (NCBI) designed an archive to accommodate these data sets. The three partners of International Nucleotides Sequencing Database Collaboration (INSDC) – NCBI, European Bioinformatics Institute (EBI) and DNA Data Bank of Japan (DDBJ), established an archive, named as Sequence Read Archive (SRA), to facilitate the scientific community an archive for NGS data sets (Shumway et al., 2009). The SRA enhance the reproducibility and facilitate the researchers for new discoveries by comparing data sets. The SRA stores raw NGS data and alignment information from different platforms that includes Roche 454, Illumina Genome Anlayzer, SOLiD system by Applied Biosystems, Heliscope by Helicos, IonTorrent by Life Technologies and Complete Genomics. The SRA database is accessible from three different collaborating organizations – NCBI[c], EBI[d] and DDBJ[e]. These three SRAs are mirror databases. The year-wise growth of SRA database is shown in Fig. 3, showing unprecedented growth of data after 2011.

---

[c]http://www.ncbi.nlm.nih.gov/Traces/sra/
[d]http://www.ebi.ac.uk/ena
[e]http://trace.ddbj.nig.ac.jp/dra/index_e.html

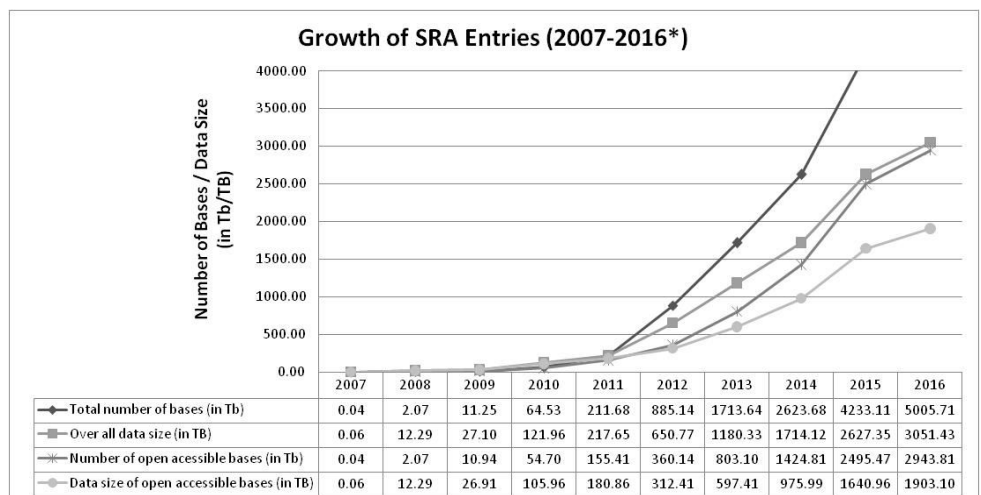

\* till May 25, 2016

Fig. 3 Year-wise growth of number of bases and its data size for SRA database

### 4.2. *RNA-Seq Atlas*

RNA-Seq Atlas (Krupp et al., 2012) is a database repository of RNA-Seq gene expression profiles and query tools. It provides free, easy and user-friendly access to RNA-Seq gene expression profiles and several tools to compare tissues as well as genes having specific expression patterns. To enhance its scope, RNA-Seq Atlas data are linked to functional and genetic databases that provides information related to corresponding genes using NCBI nucleotide database, gene signalling and pathway analysis using Kyoto Encyclopedia of Genes and Genomes (KEGG), and evaluation of gene functions using Gene Ontology (GO) database.

### 4.3. *NGS Catalog*

Another database, called NGS Catalog, has been developed by Xia and collaborators (Xia et al., 2012), that collects, curates and manages NGS data taken from published articles. This Catalog stores publication information related to NGS studies, their mutation characteristics, and mutated genesand gene fusion detected. The NGS Catalog is accessible through http://bioinfo.mc.vanderbilt.edu/NGS/. Also, this database gives a comprehensive and updated software list widely applied by NGS community, and general pipeline for the analysis of Exom-Seq, WGS and RNA-Seq data. This catalog contains over 295 records – 47 RNA-Seq datasets, 194 Exome sequencing and 54 whole genome sequencing.

### 4.4. *LANL HIV-DB Next Generation Sequence Archive*

It is a repository for next-generation lenti-virus sequence data sets containing HIV, SIV and anti-HIV/SIV antibody ultra-deep sequences. It can be accessed through http://www.hiv.lanl.gov/content/sequence/HIV/NextGenArchive/ (Archer et al., 2012).

### 4.5. *Gene Expression Omnibus*

Gene Expression Omnibus (GEO[f]) is a public repository that archives and freely distributes microarray, NGS, and other forms of high-throughput functional genomics data submitted by the research community. It supports MIAME-compliant Array as well as sequence-based data submissions. Till writing of this paper, GEO contains over 2,000 HTS platform entries, out of total ~14,500 platforms.

### 5. Computational Analysis

Various platforms of NGS technologies have made it possible to sequence huge amount of high throughput data in less time and cost. In today's era of NGS, single laboratory is able to sequence the entire human genome within a week for a few thousand dollars only (Gullapalli et. al., 2012). Production of NGS data is followed by its analysis. Careful analysis of NGS data results in important knowledge discovery and conclusions. With the advancement of time, there are uncountable efforts from researchers in developing algorithm and tool for analysis of the NGS data. Different platform need different steps and protocol in their analysis portion. To name few, some applications are CisGenome, MACS, PeakSeq, QUEST, SISSRs, CNV-Seq. A complete computational analysis workflow of NGS, starting from data preprocessing to network reconstruction and analysis, to drug discovery is shown in Fig 4. In this section, we discussed the analysis of NGS data obtained on Illumina platform.

---

[f]http://www.ncbi.nlm.nih.gov/geo/

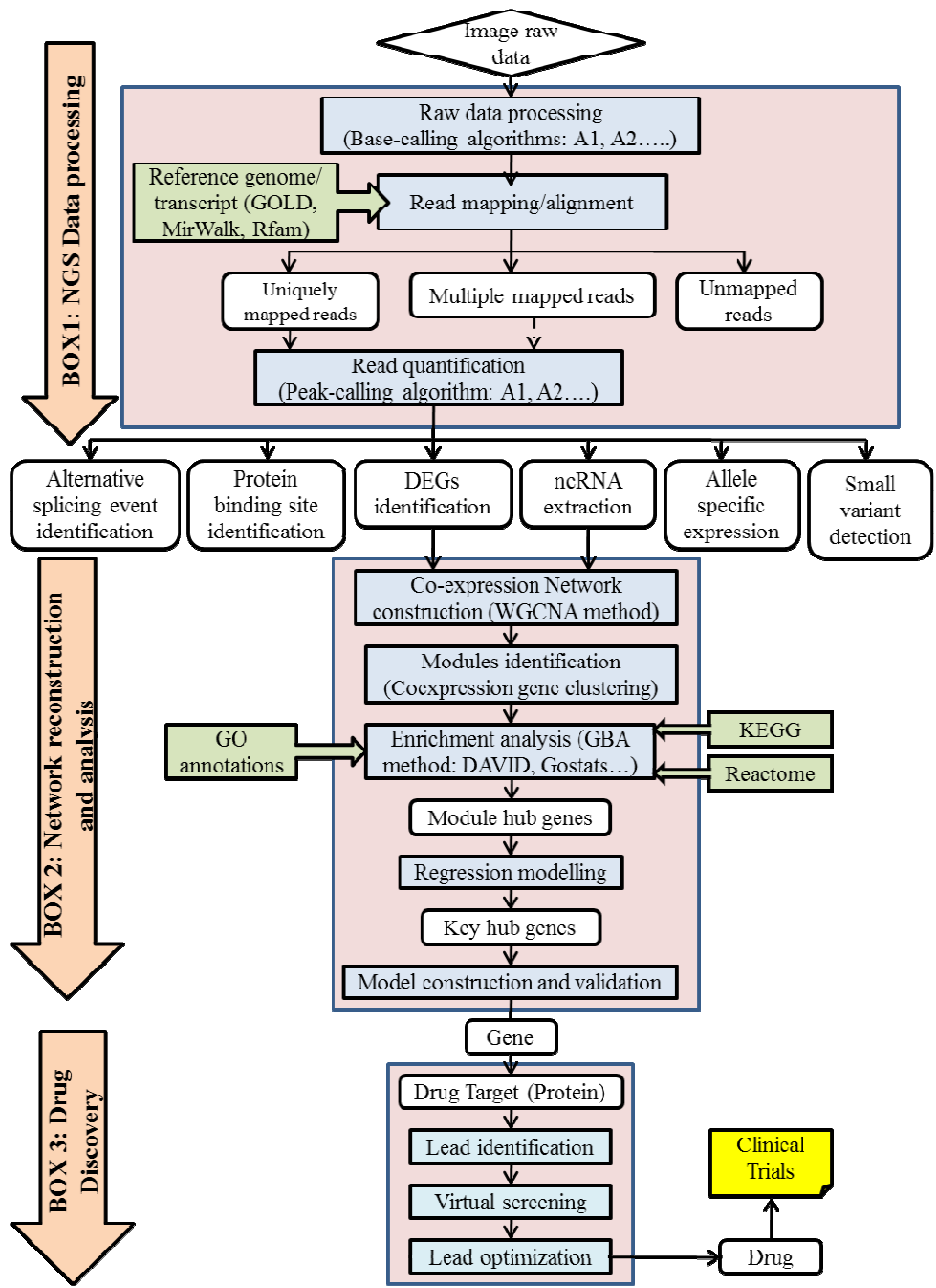

Fig. 4. The computational workflow for drug discovery starting from raw image data obtained from NGS sequencing is divided into three broad steps namely NGS data processing, Network construction and analysis, and Drug discovery. BOX 1 representing NGS data processing. The raw image data are processed into sequences by base-calling algorithms. Then reads are mapped with reference genomes/transcripts with available databases (GOLD/ MirWalk, Rfam). As a result uniquely mapped reads, multiple mapped reads and unmapped

reads can be identified. Mapped reads then undergo for read quantification with help of peak-calling algorithms which enable to identify alternative splicing event, protein binding site, differentially expressed genes, noncoding RNA, small variants in sequences and can also give insight to allele specific expression. BOX 2 depicts Network reconstruction and analysis of DEGs and ncRNA. In this step, co-expression network is constructed by WGCNA method and results into some coexpressed gene modules. Then we go for enrichment analysis of modules to identify module hub genes by Guilt-By-Association method involving GO annotations, Reactome and KEGG pathways. Then go for regression modelling to identify key hub genes from module. Then validate them by model reconstruction. Through this step we are able to get key genes for the disease. BOX 3 represents steps of Drug discovery. The identified genes can be subjected for respective protein identification (Expasy tools, etc). This protein will act as drug target having drug binding sites. Then we move towards lead identification which would give us enormous library of potent ligands that could bind protein. Then we go for virtual screening which involves selection of few potent leads based on binding affinity, etc. The most potent lead candidates are then subjected for optimization in order to maximize its druggability. Thus at last we get drug that can be further subjected to experimental and clinical trials.

### 5.1. *ChIP-Seq analysis*

Chromatin immunoprecipitation (ChIP) is commonly used in determining the Protein-DNA interactions. As discussed before, these interactions are vital in identifying the gene-expression profiles during cellular development and differentiation. ChIP-Seq is a combination of ChIP and NGS which has been widely accepted nowadays for protein-DNA interactions on a genome-wide scale. There are series of steps involved in ChIP and it is a tedious and complicated process. Eventually, after PCR amplification, there is application of massively parallel sequencing. Increasing demand and adaptation for ChIP-seq to reveal transcriptional and epigenetic regulatory mechanism in experimental laboratories round the globe has become comprehensive and sophisticated (Shin et. al, 2013). Below, we discuss simple approach of analysing the Illumina ChIP-Seq data, which helps in retrieval of useful information from the raw sequence reads.

In the first step, the sequenced data undergoes through the Image Analysis phase and Base Calling phase. The Genome Analyzer facilitates the generation of actual sequence data from the images acquired during the sequencing of this data via synthesis chemistry. In the next step, there is use of ELAND, an alignment algorithm, which aligns a set of short sequence tags against a genome. The output of this algorithm lists all the reads of the given sequence with respect to their genomic coordinates. For alignment of larger genomes (>1Mb), alignment program 'Eland_extended' is used, by calling the same in GERALD configuration file. For alignment of smaller genomes (<1Mb), alignment program 'PhagAlign' is used in places of alignment program 'Eland_extended'.

Researchers view their ChIP-seq data as a peak (signal profiles) on a genome browser. Most of them use the University of California Santa Cruz (UCSC) genome browser (Ken et. al., 2002) available over www (http://genome.ucsc.edu). UCSC genome browser has several functions. This browser is not only a web based application with standard browsing capabilities, but it also has several built features/tools to provide genomic information, evolutionary conservation, and relevant data from NIH funded genomic groups like Encyclopedia of DNA elements, commonly known as ENCODE (Raney et. al., 2010). Using UCSC genome browser, the read data which are specifically aligned to a given genome are viewed as a custom track. This track can be submitted in either BED format or Wiggle (WIG) format. BED format is a text file, which contains chromosomal start and end positions. The WIG format allows the visual examination for GC percent, probability scores and transcriptome data. Although, BED format provides better visual presentation of the data compared to the WIG format, there are limitation issues for file size, which can be uploaded on UCSC genome browser. The plus point in WIG format is its compactness, as compared to the BED format, which enables to examine larger dataset. Software requirement for ChIP-Seq analysis includes the installation of Illumina Genome Analyser Pipeline suite, which comprises of the Image analysis script, Base calling script and alignment script (GERALD), to align the reads with the reference genome.

### 5.2. *RNA-Seq analysis*

RNA-Seq basically counts the number of reads that align to one of thousands of different cDNAs, producing results similar to gene expression of microarrays. Reads are aligned to an annotated reference genome, and those aligning to exon, genes and splicing junctions are counted. Similarity, variants (CSNPs, indel) may be counted for variant analysis. The final steps are data visualization and interpretation, calculating gene- and transcript-expression and reporting differential expression. The workflow of RNA-Seq data analysis is shown in Fig. 5. In the first step, the sequencing output files are produced in compressed FASTQ format, and serve as input for next stage of the analysis. In the next stage, reads are aligned to the reference genome. The reads aligning to the exons, genes and splice junctions are counted and normalized resulting in generation of data of approximately 1.3Mb. Moreover, if desired, the cSNPs, indels can also be counted. Final stage involves the data visualization, data interpretation for the Gene- and transcript-level expression and differential expression levels. Beyond gene expression analysis, RNA-Seq can be used for i) identifying alternative splicing events, ii) allele-specific expression and iii) rate and novel transcripts, and so on.

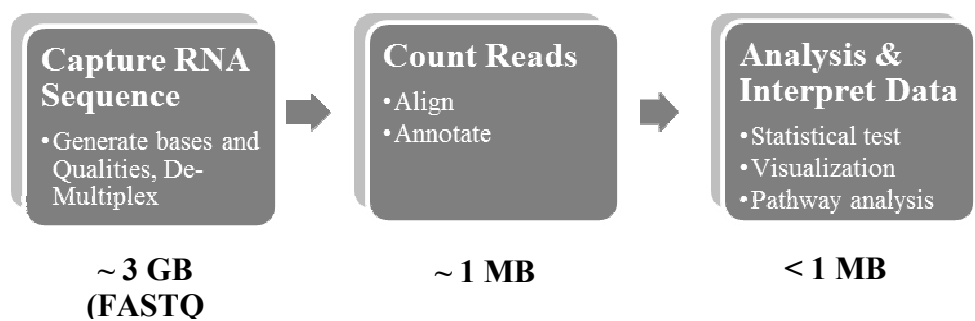

Fig. 5. Workflow of RNA-Seq data generation and analysis (Source: Data Sheet, Illumina Sequencing)

RNA-Seq is found to be vital tool for RNA profiling, discovering, and quantifying the rare, novel RNA transcripts. This sequencing technology has results similar to the gene expressions of the microarray data, but the informatics challenges and methods of analysis is different than microarray data analysis. There are two broad ways, in which the RNA-Seq analysis is different from microarray. In the case of RNA-Seq, the sequences obtained via RNA-Seq methods are generally mapped to the libraries of known exons from known transcripts. Moreover, RNA-Seq measures the number of distinct sequence reads using differential statistical tests. All these factors facilitate the RNA-Seq method in following unbiased step in analysis, which could have been possible from previous assumptions about the nature of transcriptome. This technology provides an advantage in investigating the species which have poor or missing genomic annotation. In case of microarray, the step involved in library mapping is absent and the continuous probe intensities are measured. Galaxy project is an academic initiative that allows RNA-Seq alignment, counting process, producing and uploading FASTQ files, selecting from a limited number of alignments and counting parameters.

There are several tools available for transcriptome data analysis too. Specifically, the free, open-source Tuxedo Suite comprising of Bowtie, TopHat, and Cufflinks has been widely used. Some of the commercial software providers for the analysing the RNA-Seq data are CLC Bio, GeoSpiza, GenomeQuest, and Strand Scientific Intelligence. Below we discuss the analysis protocol for Illumina based RNA-Seq data.Complex transcript data demands for highest sensitive algorithms and tools for measuring and detecting the changes in abundance and structure. The demand for improved throughput and costs requirement has enabled in development of better and sophisticated tools for RNA-Seq analysis.

### 5.3. *Non-coding RNA (ncRNA) Analysis*

It is known that most of the genes in human genome are transcribed but only a fraction of them encode for proteins. The remaining transcriptome is considered as non-coding RNA (ncRNA). Regulatory ncRNAs affect the expression and function of protein coding genes (Guffanti et al., 2014). Various studies related to cancer and ncRNA have been carried out whose goal was to discover molecular signatures and cancer biomarkers. The discovery of cancer biomarkers may reveal parameters for the early detection of cancer, its diagnosis, prognosis and prediction of response to anticancer drugs, prediction of recurrence, and identification of putative drug targets. Computational analysis of ncRNA can be comprised of three essential tasks: i) post-sequencing data analysis pipeline for ncRNa detection, classification and expression analysis; ii) a data module providing annotation information and storage for the analysis results; iii) a visualization and query system to depict and perform the functional analysis of raw data, and present results (Veneziano et al., 2015). The ncRNAs can be grouped in to various classes, each one having their own 3-dimentional folding and its own specific function, which are described in Table 3. Among all the ncRNA classes, microRNA (miRNA), long ncRNA (lncRNA) and PIWI interacting RNA (piRNA) are the most studied in cancer research (Esteller, 2011; Jorge et al., 2012).

Table 3. Different classes of ncRNA with their functions

| Non-coding RNA classes | Functions |
|---|---|
| **Micro RNA (miRNA)** | Repress gene expression |
| **Long non-coding RNAs (lncRNAs)** | Regulation of several cellular process including spigenetics, differentiation, proliferation and nuclear import |
| **Ribosomal and transport RNAs** | Essential for translation |
| **Small nucleolar RNAs (snoRNAs)** | Biogenesis and control of ribosome activity |
| **Small nuclear RNAs** | Promote splicing of pre-mRNAs |
| **Small interfering RNAs (siRNAs)** | Regulate cellular proliferation and apoptosis |
| **PIWI-Interacting RNAs (piRNAs)** | To inhibit transposable elements and DNA methylation in germ line. |
| **Circular RNAs (circRNAs)** | Untranslated, very stable, and conserved RNA molecules found in animals. |

The miRNAs consist of on average 22 nucleotides in length and mediate gene silencing by paring with specific region of mRNA to prevent its translation. The target genes of miRNA are mostly concerned with the fundamental cellular processes and aberrations in the expression of miRNAs have been rigorously studied in various cancers as they act as oncogenes (tumor suppressor genes) (Meiri et al., 2010). Also, recent studies by Matsui and colleagues (Matsui et al., 2013) suggested that miRNA may directly regulate transcription by interaction with the promoter region of divergently transcribed genes. In a study by Lui et al. (2011), they compared the expression of miRNAs having gastric cancer against normal samples. Among the 20 cancerous samples, 19 were detected as over expressed and the miR-1, miR-20a, miR-27a, miR-34, and miR-423-5p were identified as potential biomarkers for the gastric cancer diagnosis. Similarly, Wu et al., (2012) analysed samples of 42 breast cancer patients and detected over 800 circulating miRNAs. One of the earliest software tools for large-scale RNA-Seq data analysis is miRDeep (Fridlander et al., 2008), which is a stand-alone tool that can detect both known and novel miRNAs. Later, miRDeep was enhanced to miRDeep* (An et al., 2013) which has been proved to output-perform its earlier versions. Another integrated tools for the prediction and analysis of various classes of small ncRNAs on RNA-Seq data are DARIO (Fasold et al., 2011), CPSS (Zhang et al., 2012), ncPRO-seq (Chen et al., 2012), RNA-CODE (Yuan and Sun, 2013). Beside prediction of miRNA, there are several database resources for ncRNA, but the most widely used database for cancer studies are: human genome hg18 assembly (Dreszer et al., 2012), miRBase (Kozomara and Griffiths-Jones, 2010), Rfam (Gardner et al., 2011) and microRNA.org (Betel et al., 2008). Similarly, miRNA target prediction tools and its databases are TargetScan (Lewis et al., 2003), RNA Hybrid (Rehmsmeier et al., 2004), miRDB (Wang and El-Naga, 2008), DIANA-microT (Maragkakis et al., 2009), TarBase (Papadopoulos et al., 2009), miR2Disease (Jiang et al., 2009), miRecords (Xiao et al., 2009), miRWalk2.0 (Dweep and Gretz, 2015).

The lncRNAs have more than 200 nucleotides in length and most of them are still not fully characterized. Several studies reported alternations in different lncRNAs in different types of cancers (Cheng et al., 2011; Tahira et al., 2011; Yang et al., 2012).The lncRNAs, along with ncRNAs, have been identified as possible suspects in the involvement with several neurodeneration-related processes and proteins (Guffanti et al., 2014).One of the software pipelines for the detection of novel lncRNAs is called lncRScan (Sun et al,. 2013). Recently, Soreq et al., (2014) integrated full profile characterization of lncRNAs in the workflow of RNA-Seq analysis (Veneziano et al.., 2015).A comprehensive database for lncRNA, known as Noncode, became available very recently (Xie et al., 2014).

Unlike miRNAs and siRNAs which are broadly expressed in most of the cells and tissue types, the expression and functions of piRNAs are confined to germ cells and gonadal somatic cells (Aravin et al., 2007). The piRNAs, like other ncRNAs, has been found to play possible role in the biogenesis of cancer. It interacts with PIWI proteins to promote silencing of transposable elements and maintain the integrity of DNA. The circular RNAs (circRNAs), forming a covalently closed continuous loop, are untranslated, very stable and conserved RNA molecules in animals (Veneziano et al., 2015).Few circRNAs have been found as potential gene regulators, but biological functions of most of them are still unclear. Memczak et al., (2013) proposed a computational method for the detection of circRNAs in humans, mouse and C. elegans, and their finding suggests that circRNAs are often expressed in tissue or developmental-stage specific manner. Guo et al., (2014) identified and quantified human circRNAs from ENCODE Robozero RNA-Seq data (Raney et. al., 2010), and found that most of the circRNAs are expressed only in few cell types. A very recent tool, called CIRI (Guo et al., 2014), has been proved to outperform other precedent approach for the detection of circRNAs from NGS data. CIRI is an annotation independent approach deploying a de nova algorithm for the detection of novel circRNAs based on paired chiastic clipping signals combined with a filtering system to eliminate false positives.

### 5.4. *HiChIP Analysis*

Recently, a highly integrative pipeline, HiChIP (Yan et. al., 2014) has been developed for processing and systemic analysis of large amounts of ChIP-seq data. The work flow of this pipeline provides solutions to several challenges which has been reported for computational analysis of Chip-Seq/NGS data. Main functions of HiChIP are i) read quality checking; ii) read mapping and filtering; iii) peak calling and peak consistency analysis; iv) result visualizations. Moreover, this pipeline is inclusive of important modules for creating binding profiles over selected genomic features, de novo motif finding from transcription factor binding sites and functional annotation of peak associated genes. HiChIP is optimized to include options for filtering out less reliable mapped reads to reduce noise in the given experimental setup. According to the authors of HiChIP, several challenges still need to be addressed in the future. For instance, difficulties in defining the boundaries of diffuse binding sites at high resolution; identifying the direct target genes of transcriptional factor binding sites and histone modifications. It is expected that improvement in existing design and addition of new, better modular design of HiChIP will enable smooth integration of existing pipeline.

### 5.5. *ChIP-chip versus ChIP-seq*

ChIP followed by microarray hybridization is known as ChIP-chip, is similar in several aspects from recently evolved ChIP-seq. Both of these provide valuable insight to the protein-DNA interactions such as transcriptions factor bindings and histone modifications. Both ChIP-chip and ChIP-seq are widely used approaches for genome-wide identification and characterization of in vivo protein-DNA interactions and have led to several important discoveries. Though having some basic similarity, the ChIP-chip had several limitations, which was overwhelmed by ChIP-seq. Biases present in each of these platforms that can arise from both the technology and analysis methods has been experimentally reported (Ho et. al., 2011). This research reflecting the practical advantages, emphasizes the ChIP-seq to be an alternate for ChIP-chip. The Table 4 highlights the difference between these two platforms.

Table 4. Difference between ChIP-chip and ChIP-seq

| S. No. | ChIP-chip | ChIP-seq |
|---|---|---|
| 1. | Recent; first publication using this tech: 2007 | Older; first publication using this tech: 1999 |
| 2. | Data management id an issue (more than one Tb raw data is obtained usually; require huge storage space) | Data management is easy (one microarray corresponds to less than one Gb); require less storage space |
| 3. | Data analysis require powerful computers with high processing power and memory (RAM) | Computational power and memory required is comparatively lower |
| 4. | High signal to noise ratios | Low signal to noise ratios |
| 5. | More peaks | less peaks |
| 6. | Multiplexing is possible | Multiplexing is not possible |

### 6. Software Tools

Analysis of NGS data set is a challenging task that requires a systematic and intelligent approach to process it efficiently. The workflow to analyze large-scale NGS should include the processes: data quality assessment, comprehensive analysis, interpretation of results, and presentation of results in meaningful formats. There are several free as well as commercial software tools available for the analysis and visualization of NGS data sets. A list of most commonly used software tools under different task group is shown in Table 5. Most of the software tools cited is non-commercial, except few.

Table 5. List of NGS data analysis tools

| Task | Tools | URL/References |
|---|---|---|
| Integrated Tools | BaseSpace | http://www.illumina.com/informatics/research/sequencing-data-analysis-management/basespace.html |
| | CLC Genomics Workbench | http://www.clcbio.com/products/clc-genomics-workbench/ |
| | ERGO | http://www.igenbio.com/ergo |
| | Galaxy | https://wiki.galaxyproject.org/ |
| | Genomatix Genome Analyzer | http://www.genomatix.de/solutions/genomatix-genome-analyzer.html |
| | Golden Helix | http://goldenhelix.com/ |
| | JMP Genomics | http://www.jmp.com/en_us/software/jmp-genomics.html |
| | Lasergene Genomics Suite | http://www.dnastar.com/t-products-dnastar-lasergene-genomics.aspx |
| | NextGENe | http://softgenetics.com/NextGENe.html |
| | OmicsOffice for NGS (SeqSolve) | https://www.integromics.com/omicsoffice-for-ngs/ |
| | Partek Flow | http://www.partek.com/partekflow |
| | SHORE | http://1001genomes.org/software/shore.html |
| Alignment | Bowtie | http://bowtie-bio.sourceforge.net/ |
| | BFAST | http://sourceforge.net/projects/bfast/ |
| | BWA | http://bio-bwa.sourceforge.net/ |
| | SAMStat | http://samstat.sourceforge.net/ |
| | Exonerate | http://www.ebi.ac.uk/~guy/exonerate/ |
| | GenomeMapper | http://1001genomes.org/ |
| | GMAP | http://www.gene.com/share/gmap/ |
| | GNUMAP | http://dna.cs.byu.edu/gnumap/ |
| | MAQ | http://maq.sourceforge.net/ |
| | MOSAIK | https://github.com/wanpinglee/MOSAIK |
| | MUMmer | http://mummer.sourceforge.net/ |
| | Novoalign & NovoalignCS | http://www.novocraft.com/ |
| | PASS | http://pass.cribi.unipd.it/cgi-bin/pass.pl |
| | SSAHA | http://www.sanger.ac.uk/Software/analysis/SSAHA/ |
| | SOAP | http://soap.genomics.org.cn/ |
| | SWIFT | http://bibiserv.techfak.uni-bielefeld.de/swift |
| | Vmatch | http://www.vmatch.de/ |
| | ZOOM | http://www.bioinfor.com/zoom/general/overview.html |
| De Nova Assembly | AbySS | http://www.bcgsc.ca/platform/bioinfo/software/abyss |
| | ALLPATHS-LG | ftp://ftp.broadinstitute.org/pub/crd/ALLPATHS/Release-LG/ |
| | Edena | http://www.genomic.ch/edena.php |
| | MIRA4 | http://chevreux.org/projects_mira.html |
| | SOAPdenovo2 | http://soap.genomics.org.cn/soapdenovo.html |
| | SSAKE | http://www.bcgsc.ca/platform/bioinfo/software/ssake |
| | VCAKE | http://sourceforge.net/projects/vcake/ |
| | Velvet | http://www.ebi.ac.uk/~zerbino/velvet/ |
| **SNP/Indel Discovery** | GATK | https://www.broadinstitute.org/gatk/ |
| | SAM Tools | http://samtools.sourceforge.net/ |
| | SOAPsnp | http://soap.genomics.org.cn/soapsnp.html |
| | ssahaSNP | http://www.sanger.ac.uk/resources/software/ssahasnp/ |
| | PolyBayes | http://clavius.bc.edu/~marth/PolyBayes/ |
| | PyroBayes | http://bioinformatics.bc.edu/marthlab/wiki/index.php/PyroBayes |
| **Genome Annotation Browser/Viewer** | Anno-J | http://www.annoj.org/ |
| | EagleView | http://www.niehs.nih.gov/research/resources/software/biostatistics/eagleview/ |
| | Integrative Genomics Viewer | http://www.broadinstitute.org/igv/ |
| | LookSeq | http://www.sanger.ac.uk/resources/software/lookseq/ |
| | Sequence Assembly Manager | http://www.bcgsc.ca/platform/bioinfo/software/sam |
| | Staden | http://staden.sourceforge.net/ |
| | UCSC Genome Browser | http://genome.ucsc.edu/ |
| | XmatchView | http://www.bcgsc.ca/platform/bioinfo/software/xmatchview |
| **ChIP-Seq & HITS-CLIP Applications** | CisGenome | http://www.biostat.jhsph.edu/~hji/cisgenome/ |
| | CNV-Seq | http://tiger.dbs.nus.edu.sg/cnv-seq/ |
| | MACS | http://liulab.dfci.harvard.edu/MACS/ |
| | PeakSeq | http://info.gersteinlab.org/PeakSeq |
| | QUEST | http://www-hsc.usc.edu/~valouev/QuEST/QuEST.html |
| | SISSRs | http://dir.nhlbi.nih.gov/papers/lmi/epigenomes/sissrs/ |
| **Transcriptome Analysis** | E-RANGE | http://woldlab.caltech.edu/rnaseq/ |
| | G-Mo.R-Se | http://www.genoscope.cns.fr/externe/gmorse/ |
| | MapNext | http://evolution.sysu.edu.cn/english/software/mapnext.htm |
| | Partek Genomic Suite | http://www.partek.com/ |
| | RNA-Star | https://github.com/alexdobin/STAR/releases |
| | TopHat | http://ccb.jhu.edu/software/tophat |
| **Metagenomics Analysis** | MEGAN5 | http://ab.inf.uni-tuebingen.de/software/megan5/ |
| | MOCAT | http://vm-lux.embl.de/~kultima/MOCAT/ |
| | Parallel-META | http://www.computationalbioenergy.org/parallel-meta.html |
| | QIIME | http://qiime.org/ |
| | WebMGA | http://weizhonglab.ucsd.edu/metagenomic-analysis |
| **Utilities and** | FastQC | http://www.bioinformatics.babraham.ac.uk/projects/fastqc/ |
| | Picard | http://broadinstitute.github.io/picard/ |

## 7. Computational Opportunities and Challenges

NGS has dynamically enhanced the way the researchers approached in studying the DNA/RNA expressions. The technological advancements like automation, parallel processing, multiplexing has enabled production of low cost, high throughput data. Moreover, the NGS has tremendously helped in Clinical/Medical research. It is expected that proper understanding of the genomic variations and expressions will help in development of personalized medicines specific for individuals (Soon et. al., 2013). Specific strategies (as in RNA-Seq) have helped in understanding the change in expression without the need of reference genome, which was not possible in microarray data analysis. Identification of vital DNA or RNA protein interaction have been made possible using ChIP-Seq technology which have been thought to improve our epigenetic studies in the days nearby (Park, 2009). It is expected, that in the coming years, ChIP-Seq would facilitate the epigenome-wide association studies, to identify the disease mechanisms in a given population (Shin et. al., 2013). Utilization of NGS techniques have been found useful in diagnosing clinically important genetic disorders (Zhang et al, 2012). Sequencing of cancerous genome shall be exposing the altered/activated pathways, networks and other relevant information. NGS technologies not only enhanced our ability to detect genetic variations in non-coding region, but also helped us in development of enhanced statistical analysis methods in order to distinguish the true variations from sequencing errors (Kinde et. al., 2011).

NGS has made it efficient for understanding the genomics, but at the same time there are several challenges in the way. Efficient data analysis pipelines are needed to address the robustness of the NGS (Morozova & Marra, 2008). Moreover, with the passage of time, more high-throughput assays on gene expression and DNA binding activities are becoming available. This in turn, raise the requirement for systems biology approaches for identifying and inferring associative or causal relationships between genes or proteins (Shin et. al., 2013). Though NGS is a powerful tool and seems to have various applications in clinical diagnostics, but this technology stands in complex state and its successful adoption into clinical laboratory would need expertise from molecular biology and bioinformatics (Voelkerding et. al., 2009). There is a need for computational power (Data processing) as well as computational management. Storage space for huge data generated is also considered one of the challenges of NGS (Gullapalli et. al., 2012). Security of personal genomic data and ethical and political concern (for personalized genome data) have been addressed as possible challenges in the days nearby (Rizzo et. al., 2012). In addition, the cost of initial installation of NGS platforms is very much. Estimated start-up cost may rise over $100,000. Individual sequencing can cost over $1,000 per genome (Grada et. al., 2013). NGS do have false-positive and false-negative rates, because of sequencing errors and amplification biases (Soon et al, 2013), which necessitates the requirement of improved and optimized library construction methods; enhanced sequencing technologies and better filtering algorithms.

The information challenge with NGS is because of its short read lengths which makes it difficult to assemble complete genome (Schadt et al., 2010). Most of the limitations are expected to be handled by third generation sequencing (TGS) technologies such as SMRT and nano-pore technologies. However, TGS will also have modeling, data management and processing challenges leading to big data problems, demanding for HPC such as supercomputers and computer clusters. It is expected that due to availability of cloud computing based services by various companies including Amazon, Google, IBM, Joyent and Microsoft, the need for HPC and storage can be solved (Thakur et al., 2012). Further, advancement in distributed computing and availability of distributed computing architecture such as Hadoop (Shvachko et al., 2010) and Aneka (Vecchiola et al., 2009), it has become possible to handle petabyte-scale datasets distributed over a massive parallel architecture. Today, there are several companies offering cloud computing based resources for NGS and TGS data management and enable processing of large-scale raw sequence data. In comparison to NGS, there are more challenges for TGS due to scales and diversity of data it generates.

Currently, petabyte scale of raw data is available from 1000 genome projects. The availability of HPC through cloud computing services, TGS is expected to allow scanning of entire genomes, microbiomes and transcriptome within minutes with affordable cost. Scanning and mining such large-scale datasets poses several challenges for storage, processing, analysis, data transfer, access control, integration of data, and standardization of data formats.

## 8. Conclusion

The advancement of NGS technologies aided in generating genome-wide sequence data in a couple of days, and this huge amount of data led to numerous important discoveries. However, this technology needs specific data analysis algorithms which are written for specific task to be performed. The algorithms in group, act as a tool in analyzing the NGS data. Analysis of NGS data unravels important clues in quest for the treatment of various life-threatening diseases; improved crop varieties and other related scientific problems related to human welfare. This review covers basic background of NGS technologies. We discussed here the possible applications and computational approaches for analysis of various types of NGS data, including ChIP-Seq analysis, RNA-Seq analysis, Non-coding RNAs identification and HiChIP analysis. Importance of ChIP-Seq over ChIP-chip is discussed in brief. Moreover, we have also provided details of available database archives for NGS data and also exhaustive list of algorithms/software to perform various tasks during NGS data analysis. It is expected that this review covers present computational opportunities and also the challenges in the analysis and interpretation of NGS data.

**List of Abbreviations**

Capillary Electrophoreses (CE); Chromatin Immunoprecipitation Sequencing (ChIP-Seq); Free and Open Source Software (FOSS); Gene Expression Omnibus (GEO); Gene Ontology (GO); High Performance Computing (HPC); High-throughput sequencing (HTS); Human Genome Project (HGP); Kyoto Encyclopedia of Genes and Genomes (KEGG); Massively Parallel Signature Sequencing (MPSS); Mega base pair (Mbp); National Institute of Health (NIH); Next-Generation Sequencing (NGS); Non-coding RNA (ncRNA); Personal Genome Machine (PGM); Polymerase Chain Reaction (PCR); Sequence Read Archive (SRA); Single Molecule Real-Time (SMRT); Single Molecule Sequencing (SMS); Single Nucleotide Polymorphisms (SNPs); Third Generation Sequencing (TGS); Whole Genome Sequencing (WGS).

**Competing interests**

The authors declare that they have no competing interests.

2727

31